\documentclass{PoS}

\usepackage{amsmath}
\usepackage{amsthm}

\title{The $\mathbb{Z}_3$ model with the density of states method}

\ShortTitle{The $\mathbb{Z}_3$ model with the density of states method}

\author{\speaker{Ydalia Delgado Mercado}\\
        Institut f\"ur Physik, Karl-Franzens Universit\"at Graz, 8010 Graz, Austria\\
        E-mail: \email{ydelgado83@gmail.com}}
  
\author{Pascal T\"orek\\
        Institut f\"ur Physik, Karl-Franzens Universit\"at Graz, 8010 Graz, Austria\\
        E-mail: \email{pascal.toerek@edu.uni-graz.at}}  
        
\author{Christof Gattringer\\
        Institut f\"ur Physik, Karl-Franzens Universit\"at Graz, 8010 Graz, Austria\\
        E-mail: \email{christof.gattringer@uni-graz.at}}

\abstract{In this contribution we apply a new variant of the density of states method to the $\mathbb{Z}_3$ spin model 
at finite density. We use restricted expectation values evaluated with 
Monte Carlo simulations and study their dependence on a control parameter $\lambda$. We show that a sequence of one-parameter fits
to the Monte-Carlo data as a function of $\lambda$ is sufficient to completely determine the density of states. We expect that this method 
has smaller statistical errors than other approaches since all generated Monte Carlo data are used in the determination of the density. We compare 
results for magnetization and susceptibility to a reference simulation in the dual representation of the $\mathbb{Z}_3$ spin model  
and find good agreement for a wide range of 
parameters.  }

\FullConference{The 32nd International Symposium on Lattice Field Theory\\
		 23-28 June, 2014\\
		 Columbia University New York, NY}

\begin{document}

\section{Introduction}
Lattice QCD at finite chemical potential $\mu$ is notoriously difficult for Monte Carlo simulations since for $\mu > 0$
the fermion determinant and thus the
effective fermion action are complex. The corresponding Boltzmann factor can no longer be interpreted as a probability and thus is not suitable 
for importance sampling.  Various methods to overcome this so-called complex action problem have been explored, such as reweighting, complex
Langevin, various expansions around $\mu = 0$ and rewriting partition sums to different degrees of freedom (dual variables). 

\hspace*{8mm}
Another possible approach is the density of states (DoS) method. Here the complex action problem manifests itself in the fact that for the evaluation 
of observables the density of states is multiplied with a rapidly oscillating factor, such that for reliable results the density of states has to be 
determined with very high accuracy. Recently an interesting variant of the DoS method has been proposed  \cite{dos,Z3dos} for
lattice field theories where a restricted 
Monte Carlo update is used to obtain exponential error reduction in the determination of the density of states. 

\hspace*{8mm}
In this contribution we apply the density of states method to the $\mathbb{Z}_3$ spin model, where a lattice simulation with dual variables 
is available \cite{Z3worm} and provides reference data to test the reliability and accuracy of the DoS approach. The model has been studied 
also using fugacity and Taylor expansion \cite{Z3fugacity} and also a previous study with DoS techniques has been published in \cite{Z3dos}. 

\hspace*{8mm}
The variant of the DoS approach we present here is 
slightly different from the one in \cite{Z3dos}: We use a similar ansatz to parameterize the 
density of states $\rho$, but determine the coefficients of this parameterization in a 
different way, which we refer to as the functional fit approach (FFA):
In the restricted Monte Carlo simulation we study the dependence on a free control parameter $\lambda$ and fit the known functional 
form as a function of $\lambda$ to the Monte Carlo data. We show that a sequence of one parameter fits 
is sufficient to obtain all parameters of the density. We expect that 
this approach has smaller statistical errors and is numerically more stable than root finding or iterative techniques, since all Monte Carlo data
generated for different values of $\lambda$ are used to determine $\rho$.

\section{Definition of the model and the density of states}
 
The $\mathbb{Z}_3$ spin model in an external field with strength $\kappa$, a chemical potential $\mu$ and a temperature 
parameter $\tau$ is described by the action
 
\begin{equation}
S[P]  =  \sum_x\left[\tau\sum_{\nu=1}^3 \big(P_x^\star P_{x+\hat{\nu}} + c.c. -2\big) + \kappa  e^\mu (P_x-1)  + 
\kappa e^{-\mu} (P_x^\star-1) \right],
\label{action}
\end{equation}
where the dynamical degrees of freedom (spins) are $P_x\in \mathbb{Z}_3 = \{1,e^{i2\pi /3},e^{-i2\pi /3}\}$, living on the sites
$x$ of a 3-dimensional lattice with periodic boundary conditions. The action is normalized such that $S[P]=0$ if $P_x=1\,\forall x$. 
The partition function is obtained by summing the Boltzmann factor
over all configurations $\{P\}$, i.e.,  $Z = \sum_{\{P\}} e^{S[P]}$. It is obvious that for finite chemical potential, $\mu \neq 0$, 
the action (\ref{action}) has a non-zero imaginary part and the model has a complex action problem.

\hspace*{8mm}
For a more convenient notation we introduce abbreviations for the total numbers of spins pointing in each of the three directions as
$N_0[P] \; = \; \sum_x \delta\left(P_x,1\right)$ and $N_\pm[P] \; = \; \sum\limits_x\delta\left(P_x,e^{\pm i 2\pi/3}\right)$,
and note that obviously $N_0 + N_+ + N_- = V$, where $V$ denotes the lattice volume, i.e., the total number of sites of the lattice.
Using these we can rewrite the action to the form  
\begin{eqnarray}
S[P] & = & S_R[P] \, + \, iS_I[P] \; , \label{eq1.2} \\
S_R[P] & = & \tau \sum_x \sum_{\nu=1}^3\big(P_x^\star P_{x+\hat{\nu}} + c.c.-2\big) \; + \; \kappa \, 3(N_0[P] - V) \cosh \mu  \; ,
\nonumber \\
S_I[P] & = & \kappa \, \sqrt{3} \, \sinh \mu \, \Delta N[P] \; , 
\nonumber
\end{eqnarray}
where we have defined
$ \Delta N[P] = N_+[P] - N_-[P] \; \in \; \{-V,-V+1,\dots,V\} $.

\hspace*{8mm}
Exploring the properties of $S_R[P]$ and $S_I[P]$ under complex conjugation of the spin variables we find for the partition sum
\begin{equation}
Z \; = \; \sum_{\{P\}} e^{S[P]} \;  = \; \sum_{\{P\}} e^{S_R[P]} \, \cos\left(S_I[P]\right) \; = \; 
\sum\limits_{\{P\}} e^{S_R[P]}\cos\left(\kappa \sqrt{3} \sinh \mu \; \Delta N[P] \right) \; .
\label{partitionsum}
\end{equation}
We now define a weighted density of states which is a function of $d \equiv \Delta N$,
\begin{equation}
\rho(d) \; = \; \sum_{\{P\}}e^{S_R[P]} \, \delta\left(d - \Delta N[P]\right) \; ,
\label{rhodef}
\end{equation}
which can be shown to be an even function of $d$, i.e.,
$\rho(-d) \; = \; \rho(d)$.
Using the density of states, the partition sum and expectation values of observables which are a function $O(\Delta N)$ of $\Delta N$ 
can be written as
\begin{eqnarray}
Z  & = & \sum_{d\, =\, -V}^{V}\rho(d) \, \cos\left( \kappa \sqrt{3} \sinh\mu \; d  \right) \; ,
\label{zfinal}
\\
\langle O \rangle  & = & \frac{1}{Z} \sum_{d \, = \, -V}^{V}  \rho(d) \, \Big[ \cos\left( \!\kappa \sqrt{3} \sinh \mu  \, d  \!\right) O_E(d) \; + \; i 
\sin\left( \!\kappa \sqrt{3} \sinh \mu  \, d  \! \right) O_O(d) \Big] \, ,
\label{vev}
\end{eqnarray}
where $O_E$ and $O_O$ denote the even and odd parts of $O(\Delta N)$.

\hspace*{8mm}
The partition sum and expectation values are obtained by reweighting the density $\rho(d)$ with the factors
$\cos\left(\kappa \sqrt{3} \sinh(\mu) \, d \right)$ and $\sin\left(\kappa \sqrt{3} \sinh(\mu) \, d \right)$. 
While the density $\rho(d)$ is strictly positive, these factors are oscillating with $d$
and the frequency of oscillation increases exponentially with  $\mu$. Thus for larger values of 
$\mu$ the density $\rho(d)$ has to be computed very accurately. This is how the complex action problem manifests 
itself in the density of states approach. 
    
\section{Computing the density of states}
For the numerical computation we parameterize the density of states $\rho(d)$, with $d \in [-V,V]$, as
\begin{equation}
\rho(d) \; = \; \prod_{i=0}^{|d|} e^{-a_i} \; = \; \exp \left( -\sum_{i=0}^{|d|} a_i \right) \; ,
\label{rhoparam}
\end{equation}
with real parameters $a_i$. Note that this parameterization is exact in the sense that it contains $V+1$ parameters, 
precisely the number of independent degrees of freedom $\rho(d)$ has (remember that $\rho(d)$ is an even function). 
We also remark, that an overall normalization of $\rho(d)$ can be chosen freely, since it 
cancels in the expectation values (\ref{vev}). Here we choose the normalization $\rho(0) = 1$, which corresponds to 
setting $a_0 = 0$. 
  
\hspace*{8mm}
For the calculation of the coefficients $a_i$ we define restricted expectation values $\langle \langle O \rangle \rangle_n(\lambda)$,
$n = 0,1, ... \, V-1$,
which depend on a free parameter $\lambda$,
\begin{equation}
Z_n(\lambda) \, = \, \sum_{\{ P \}} 
\theta_n\big(\Delta N[P]\big) \, e^{S_R[P]} \,  e^{\, \lambda \, \Delta N[P]} \; , \; 
\langle\langle O \rangle\rangle_n (\lambda) \, =\,  \frac{1}{Z_n(\lambda)} \sum_{\{ P \}} 
\theta_n\big(\Delta N[P]\big) \, e^{S_R[P]} \, e^{\, \lambda \, \Delta N[P]} \, O( \Delta N[P] ) \;  .
\label{vevrestricted}
\end{equation}
Here we defined
\begin{equation}
\theta_0(d) \; = \; \left\{ \begin{array}{cc} 
1 \; , \; & \mbox{for} \; d =  0, 1 \\
0 \; , \; & \mbox{otherwise} 
\end{array} \right., \quad
\mbox{and} \; \theta_n(d) \; = \; \left\{ \begin{array}{cc} 
1 \; , \; & \mbox{for} \;  |d -n| \leq 1 \\
0 \; , \; & \mbox{otherwise} 
\end{array} \right. \,  \mbox{for} \; n = 1,2, ... \, V - 1\; .
\end{equation}
In the restricted expectation values (\ref{vevrestricted}) only real and positive weight factors appear, such that they 
can be evaluated with a restricted Monte Carlo strategy which we will 
discuss below. In particular we are here interested in the observable $O = \Delta N$.

\hspace*{8mm}
We can now also use the density of states $\rho(d)$ in the form of (\ref{rhoparam}) to evaluate the restricted  
expectation values $\langle\langle \Delta N \rangle\rangle_n (\lambda)$. A straightforward calculation gives  
\begin{equation}
\langle\langle \Delta N \rangle\rangle_0 (\lambda) \; = \;  
\frac{e^{\lambda - a_1}}
{e^{\lambda - a_1} + 1}  \; , \;
\langle\langle \Delta N \rangle\rangle_n (\lambda) - n \hspace{3mm}  =  \frac{ e^{2 \lambda - a_n - a_{n+1} } - 1}
{ e^{2 \lambda - a_n - a_{n+1} } +
e^{ \lambda - a_n } + 1 } \; \; , \; \; n = 1, ... V \! - \! 1 \; .
\label{vevn} 
\end{equation}
The right hand sides are simple functions of $\lambda$: They are monotonically increasing (their derivatives 
with respect to $\lambda$ are easily shown to be positive) and for $n \geq 1$ they have a single zero ($\lim_{\lambda \rightarrow -\infty}$ 
is negative, $\lim_{\lambda \rightarrow + \infty}$ is positive). Examples for different $n$ are shown in Fig.~\ref{fig_coeffs} below.

\hspace*{8mm}
Using Monte Carlo simulations we can evaluate  
$\langle\langle \Delta N \rangle\rangle_n (\lambda) - n$ for different values $\lambda_i, \,  i = 1,2, ... \, N_\lambda$ (typically 
$N_\lambda = {\cal O}(10)$) and fit the results according to the right hand sides of (\ref{vevn}). The one-parameter fit 
for $\langle\langle \Delta N \rangle\rangle_0 (\lambda)$ determines the first non-trivial coefficient $a_1$ 
(remember that we chose the normalization $a_0 = 0$). The fit value for $a_1$ can then be inserted in the right hand side 
of (\ref{vevn}) for $n=1$ such that with 
another one-parameter fit of  $\langle\langle \Delta N \rangle\rangle_1 (\lambda) - 1$ we can determine $a_2$, which in turn is then 
inserted in the fit function for $\langle\langle \Delta N \rangle\rangle_2 (\lambda) - 2$, which gives $a_3$ from a one-parameter fit, 
et cetera. Using this sequence of fits we can determine all coefficients $a_i$ from fits of the Monte Carlo data with 
simple functions depending on a single parameter (compare Fig.~\ref{fig_coeffs}).

\section{\large Restricted Monte Carlo}

The variant of the density of states method described here is based on fitting the Monte Carlo data for the restricted 
expectation values $\langle \langle \Delta N \rangle \rangle_n(\lambda)$ as defined in (\ref{vevrestricted}). For this purpose we 
first need to generate initial configurations $P$ of the spin variables, such that the constraint $\Delta N [P] \in \{ n- 1, n, n +1\}$ 
is obeyed (for $n > 0$).   
These configurations can easily constructed by hand, but of course need to be 
equilibrated before taking measurements (we use $10^6$ equilibration sweeps for the data we show).
Once the initial configurations that obey the constraints are generated, a slightly modified conventional Monte Carlo update 
can be used, with the additional restriction that proposed trial configurations which violate the constraint 
$\Delta N [P] \in \{ n- 1, n, n +1\}$ are rejected. The acceptance rate is very good throughout and only for $n$ very close to the
maximum value of $n=V$ (i.e., the cases $n = V, V -1, V - 2$) we observe a strong drop in the acceptance rate. In principle it is easy to 
compute $\rho(d)$ for these largest values of $d$ exactly with a low temperature expansion. However, since for the values of $d$ 
where the quality of the Monte Carlo data decreases $\rho(d)$ is already very small, we simply use the data as we obtain them from
the simulation. 

\begin{figure}[t]
\centering
\hspace*{-5mm}\includegraphics[width=0.7\textwidth]{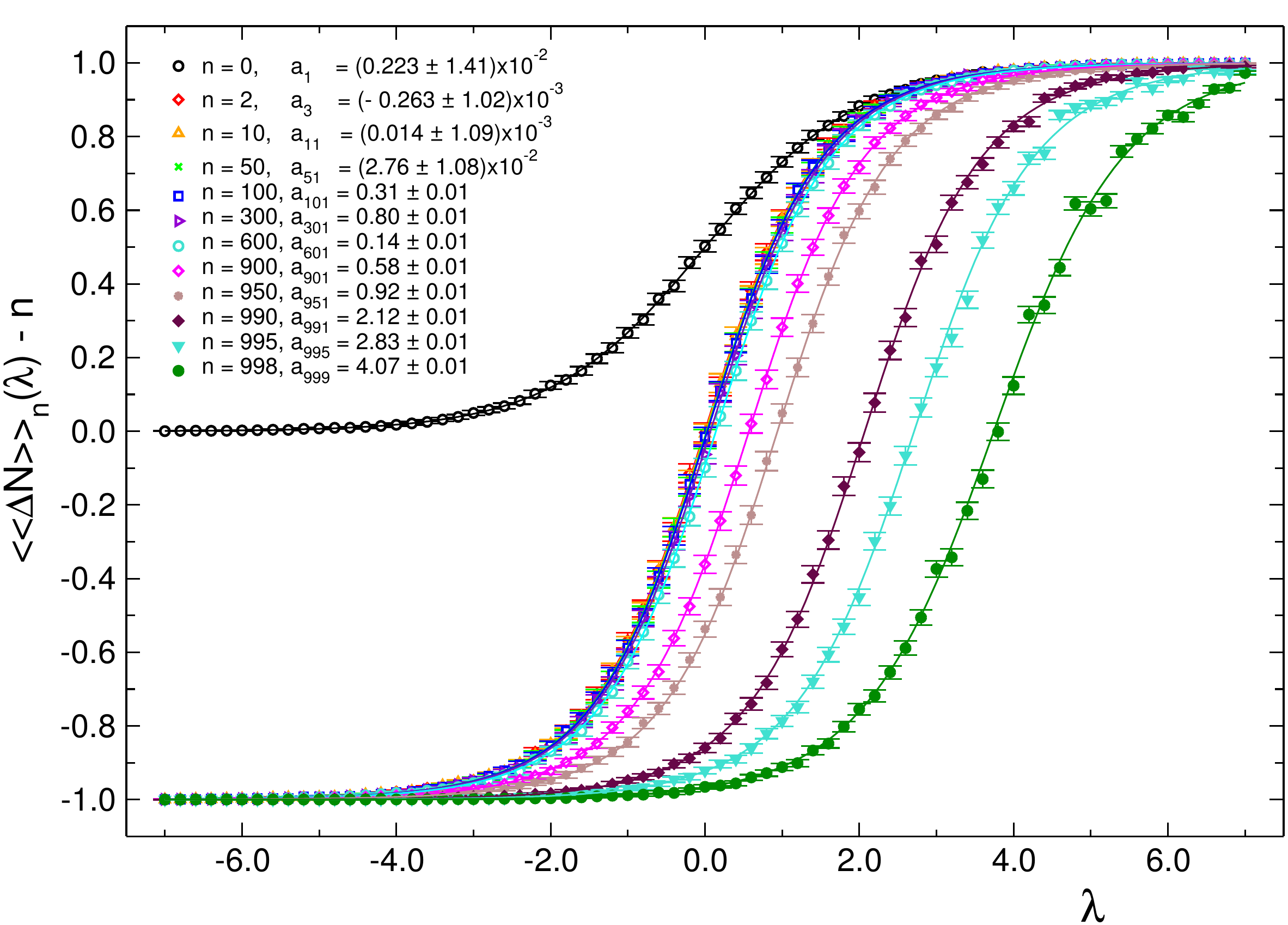}
\caption{Monte Carlo results (symbols) for $\langle \langle \Delta N \rangle \rangle_n(\lambda)\!-\!n$ for 
$n = 0$, 2, 10, 50, 100, 300, 600, 900, 950, 990, 995 and 998 as a function of $\lambda$. The data are for  the parameters $\tau = 0.16$, 
$\kappa = 0.01$ and $\mu = 1.0$. The full curves are the fits with the functions on the rhs.\ of (3.4).
The resulting values for the corresponding coefficients $a_n$ are given in the legend.
\label{fig_coeffs}}	
\end{figure}

\hspace*{8mm}
In Fig.~\ref{fig_coeffs} we show the Monte Carlo results (symbols) for 
$\langle \langle \Delta N \rangle \rangle_n(\lambda) - n$ with $n = 0$, 2, 10, 50, 100, 300, 600, 900, 950, 990, 995 and 998  
for several values of
$\lambda$ in the interval $[-7,7]$. The data were generated on $10^3$ lattices for the parameter values $\tau = 0.16$, 
$\kappa = 0.01$ and $\mu = 1.0$. After $10^6$ equilibration sweeps we sample $10^5$ configurations separated by 100 sweeps 
for each data point, and the errors we show are the statistical errors. 
The figure demonstrates that the Monte Carlo data show the expected simple behavior as a function of $\lambda$ and can easily 
be fit (we use a standard $\chi^2$ procedure) with the functions given in the right hand sides of (\ref{vevn}).
The results of the fits are shown in Fig.~\ref{fig_coeffs} as full curves and it is obvious that they perfectly describe the numerical data.

\begin{figure}[t]
\centering
\hspace*{-3mm}
\includegraphics[width=0.4\textwidth]{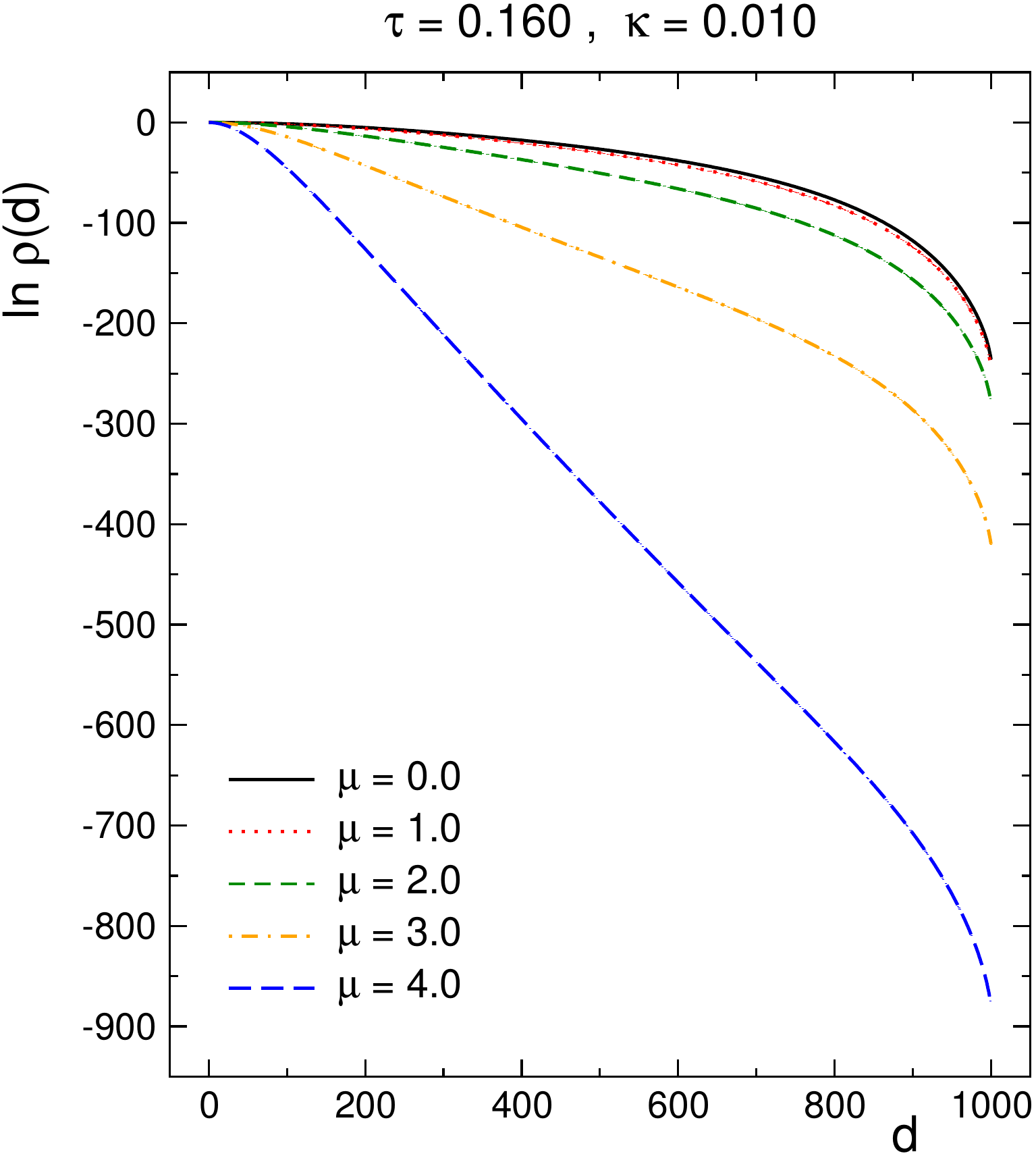}
\includegraphics[width=0.39\textwidth]{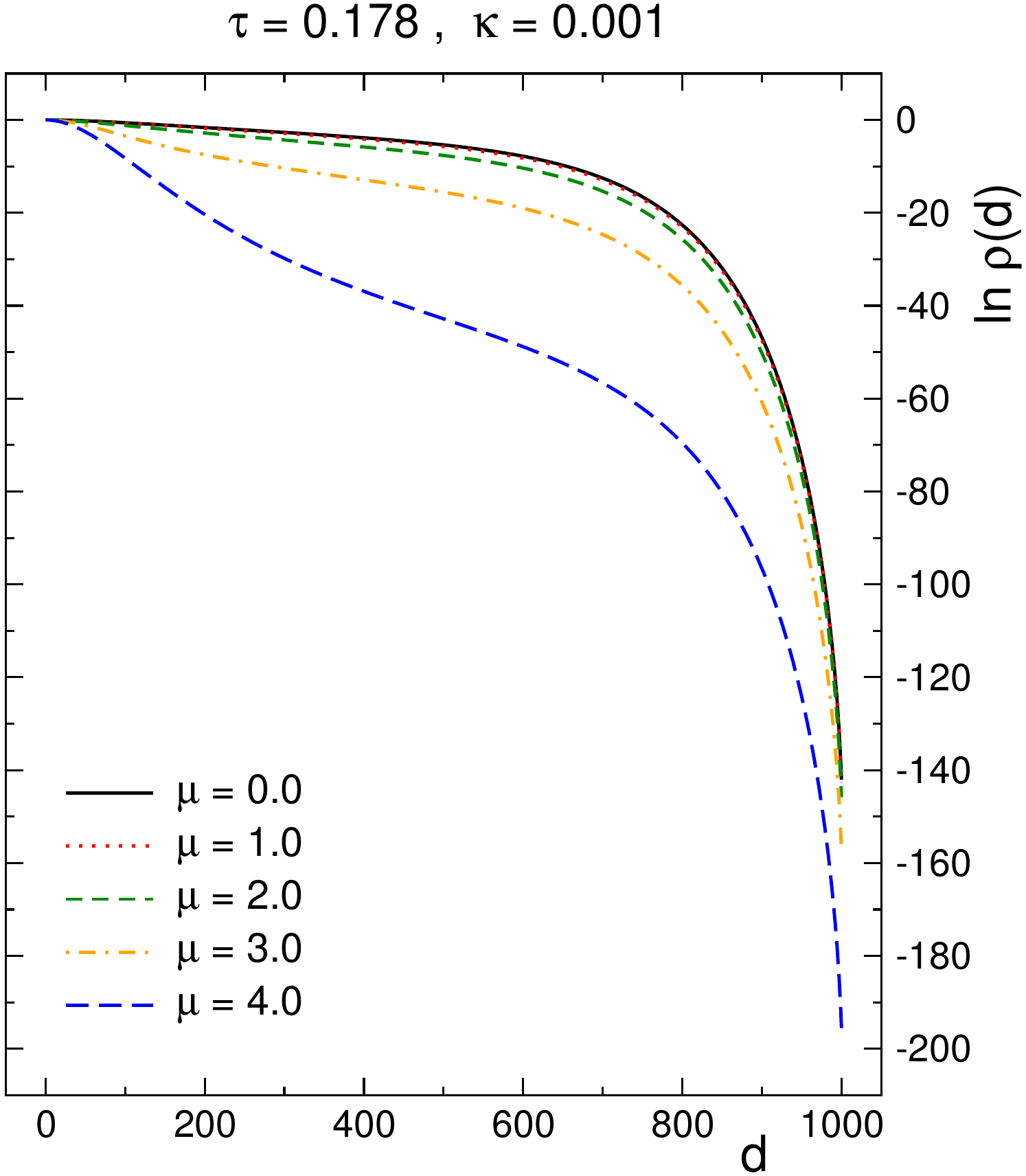}
\caption{Results for the logarithm of $\rho(d)$ as a function of $d$ from a $10^3$ lattice for different values of $\mu$. 
The data we show in the lhs.\ plot are for 
$\tau = 0.16$, $\kappa = 0.01$. On the rhs.\ we use $\tau = 0.178$ and $\kappa = 0.001$. The error bars are smaller than the line-width. Note the different vertical scales for the two plots.}
\label{rhoexample}	
\end{figure}

\hspace*{8mm}
Once the coefficients $a_i$ are determined from the fits, we can build up the density of states $\rho(d)$ as given in 
(\ref{rhoparam}).  Results for the density $\rho(d)$ at different values of $\mu$ are presented in Fig.~\ref{rhoexample}.  
The data we show in the lhs.\ plot are for 
$\tau = 0.16$, $\kappa = 0.01$, while in the rhs.~plot $\tau = 0.178$, $\kappa = 0.001$ were used. It is remarkable that the range 
of the values for $\rho(d)$ strongly depends on the parameters, including also $\mu$. This is due to the fact that the density 
we use here also includes the $\mu$-dependent Boltzmann factor $e^{S_R}$.

\section{\large Results for physical observables}

\begin{figure}[t!]
\centering
\hspace*{-2mm}
\includegraphics[width=0.43\textwidth]{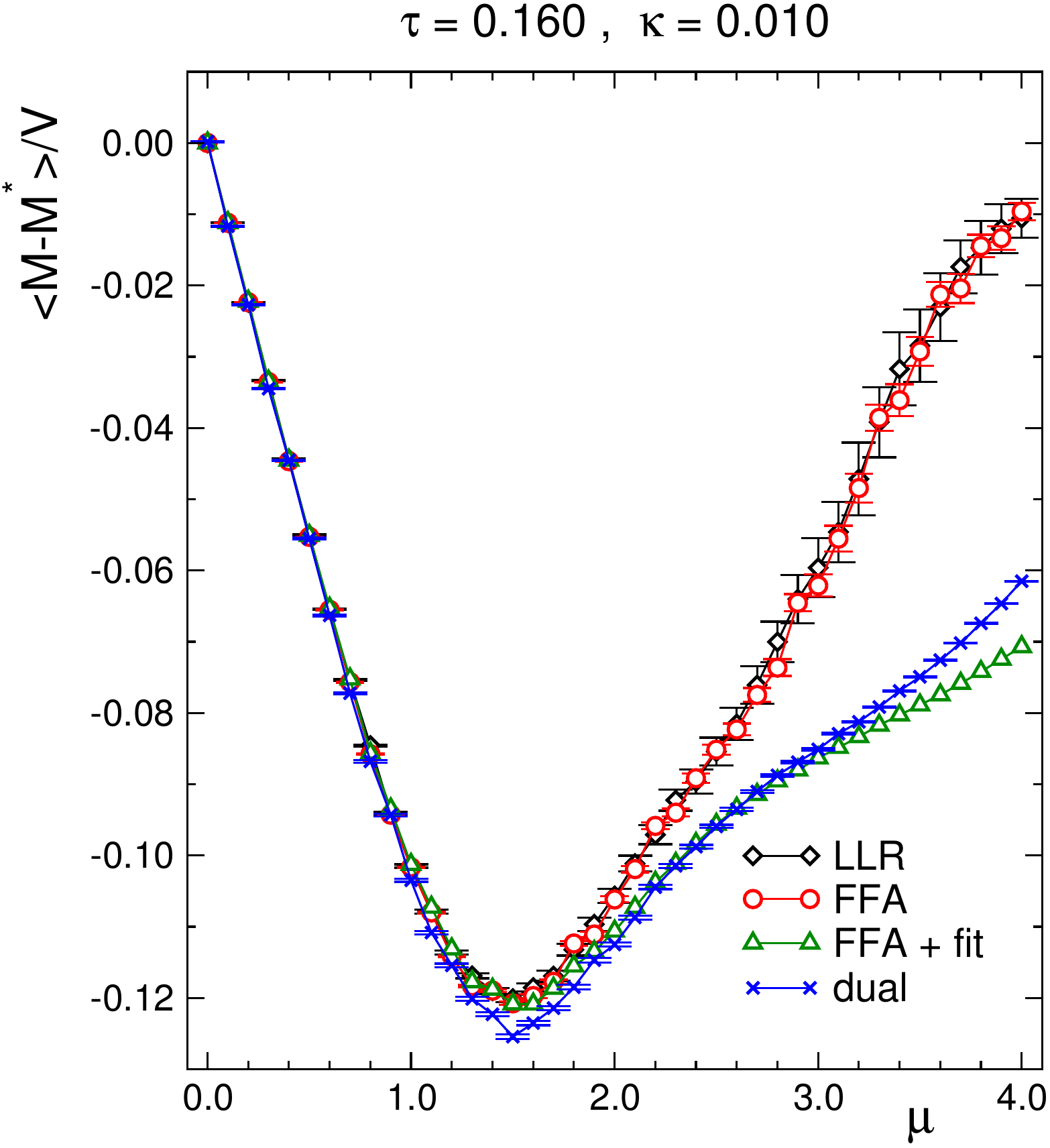}
\hspace{-1mm}
\includegraphics[width=0.43\textwidth]{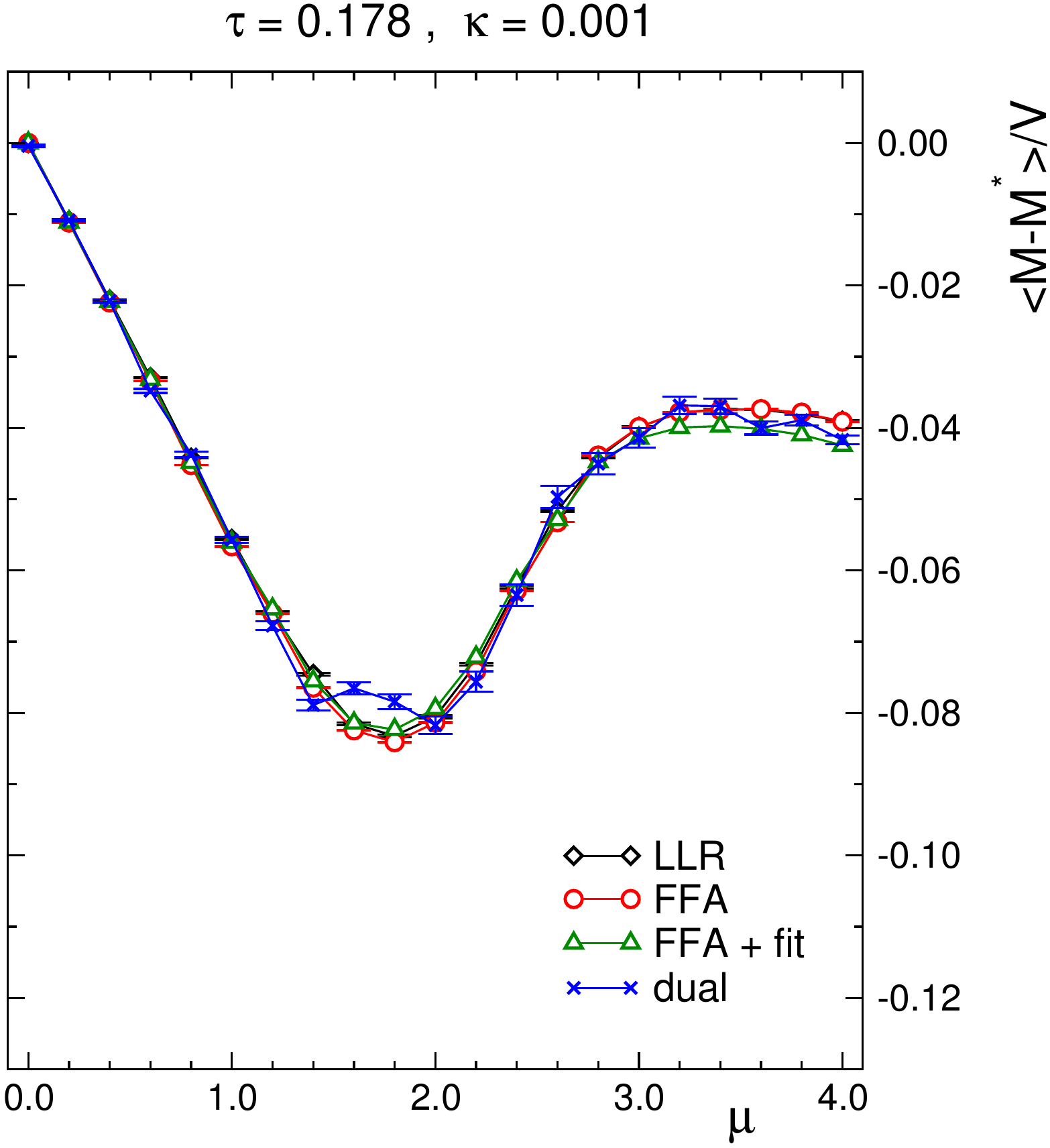}
\vspace{5mm}
\hspace*{0.1mm}
\includegraphics[width=0.405\textwidth]{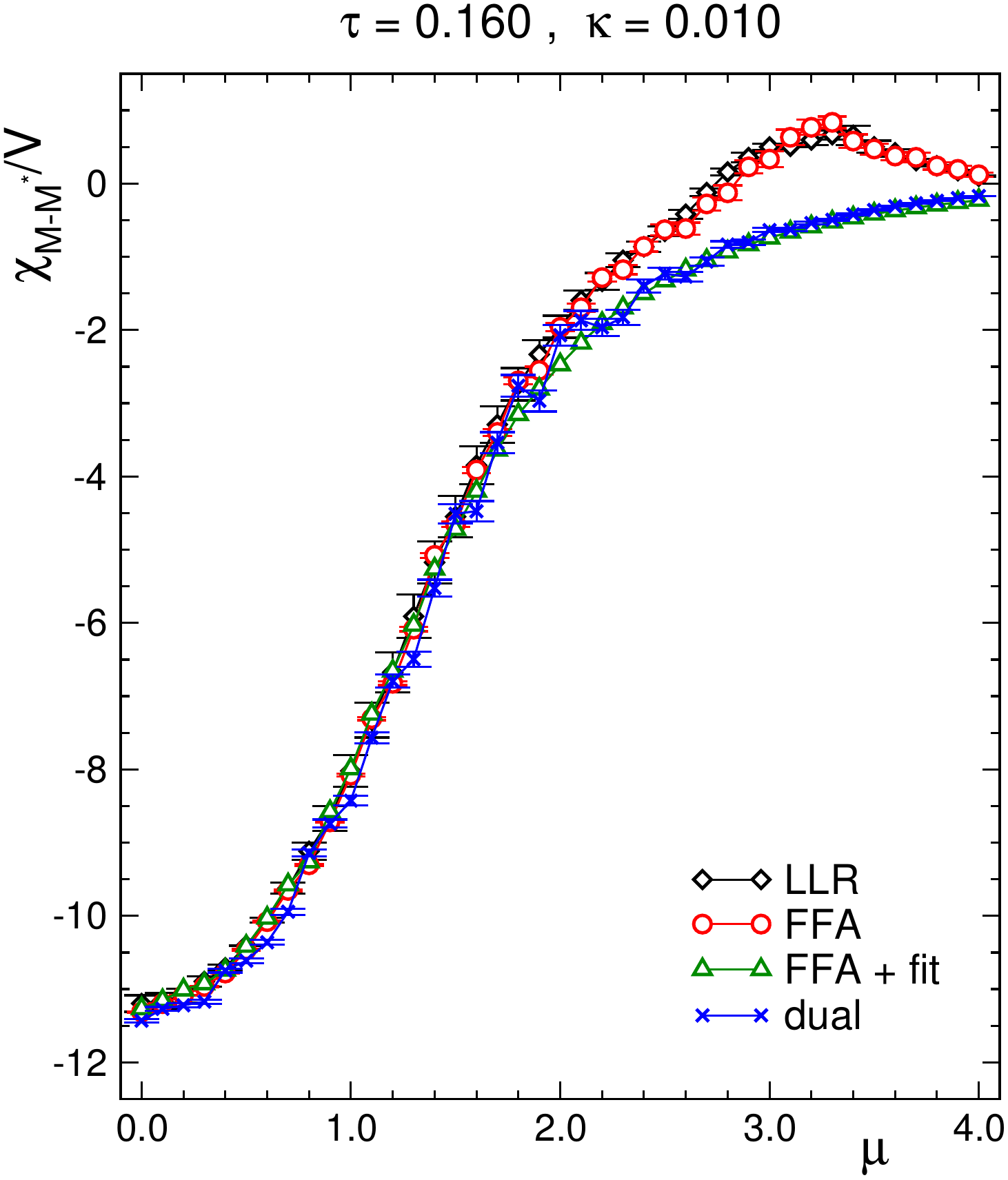}
\hspace{-1.2mm}
\includegraphics[width=0.416\textwidth]{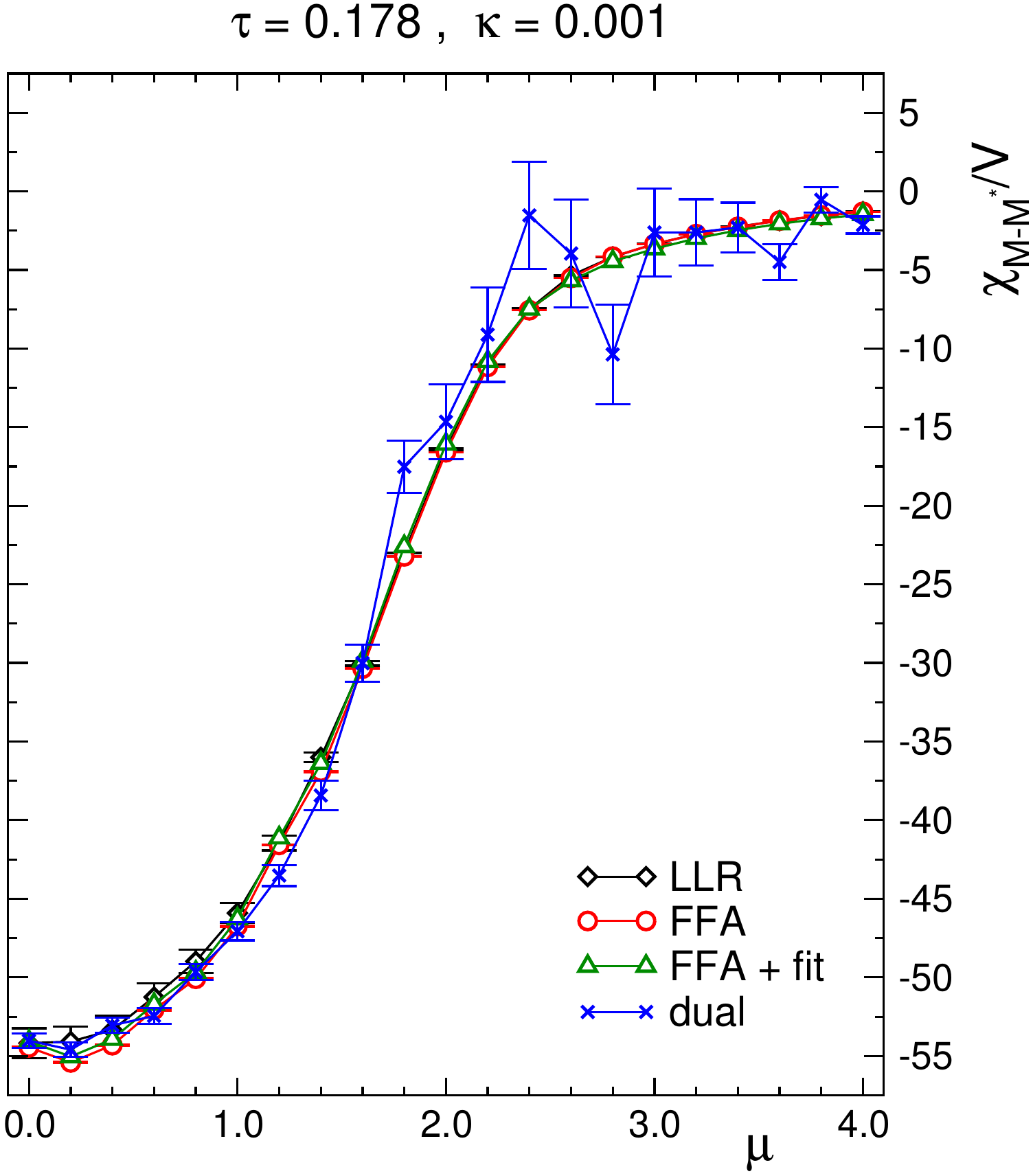}
\caption{Results for the physical observables $M-M^*$ (top row of plots) and $\chi_{M-M^*}$ (bottom) on $10^3$ lattices
as a function of $\mu$. We use  
two sets of parameters, $\tau = 0.16$, $\kappa = 0.01$ on the lhs., and $\tau = 0.178$ and $\kappa = 0.001$ on the rhs. 
We show results from the LLR algorithm, the FFA algorithm, the FFA algorithm combined with a fit of $\rho(d)$ and for comparison also the results from a simulation in the dual representation.}
\label{observables}	
\end{figure}

Having determined the density $\rho(d)$ we can finalize the calculation and evaluate expectation values of observables
using (\ref{vev}). We consider 
$\langle M - M^* \rangle  \quad$ and $\quad \chi_{M - M^*}  =   
\langle (M - M^*)^2 \rangle - \langle M - M^* \rangle^2$,
where $M  =  \sum_x P_x$.
In Fig.~\ref{observables} we show the results from the FFA for $\langle M - M^* \rangle$ and $\chi_{M - M^*}$ as a function of $\mu$ (red circles). 
We display results for two sets of parameters, $\tau = 0.16$, $\kappa = 0.01$ on the lhs., and $\tau = 0.178$, $\kappa = 0.001$ (rhs.). 
The data are for $10^3$ lattices and the statistics is as described in the previous section. 
The results are compared to the observables evaluated with the LLR approach for the determination of the density (density parameterized with 
$\Delta d = 4$, same statistics as the FFA, black diamonds). It is obvious, that the results from the FFA and the LLR approach agree well, but that 
the FFA data have smaller statistical errors as expected.  

\hspace*{8mm}
As reference data in Fig.~\ref{observables} we also show the results from a dual simulation \cite{Z3worm}
($10^6$ measurements, crosses). 
For the $\tau = 0.178, \kappa = 0.001$ data (rhs.) the FFA and LLR results agree well with the dual results for all values of $\mu$ 
we show. However, for $\tau = 0.16$, $\kappa = 0.01$ (lhs.) the dual and the DoS data disagree for $\mu$ larger than 2.
This discrepancy can be attributed to the fact, that for this parameter set the sign problem is much harder than for $\tau = 0.178, \kappa = 0.001$
(compare \cite{Z3fugacity}). In addition we also show the results when the density
$\rho(d)$ is replaced by a fit of $\rho(d)$ with a finite polynomial in $d^2$ \cite{Z3dos}. 
This reduces the impact of local fluctuations and increases the range of 
$\mu$ where the DoS results agree with the dual simulation.

\vskip3mm
\hspace*{8mm}
To summarize: Our assessment shows that the density of states method based on restricted Monte Carlo simulations
\cite{dos,Z3dos} is certainly a competitive and rather generally applicable method. To obtain a maximal range of values for the chemical potential 
$\mu$, the density has to be computed as precisely as possible. In the variant we discuss here this is done by fitting the restricted Monte Carlo 
data to the dependence on the control parameter $\lambda$ and in this way making optimal use of all computed Monte Carlo data. 
Technical improvements such as these will contribute to the further refinement of the density of states method.

 \vspace{0.3cm}
{\bf Acknowledgments:}
We thank Biagio Lucini and Kurt Langfeld for fruitful discussions. Y. Delgado Mercado was partly 
funded by the FWF DK W1203 ``{\sl Hadrons in Vacuum, Nuclei and Stars}''. Furthermore this project is supported by DFG TR55, 
''Hadron Properties from Lattice QCD'' and by the Austrian Science Fund FWF Grant. Nr. I 1452-N27.

\end{document}